\documentclass[preprint,amsmath,amssymb,showkeys]{revtex4-1}
\usepackage{graphicx}
\usepackage{xcolor}
\usepackage{enumerate}
\usepackage{longtable}

\begin{document}

\title{A worldwide model for boundaries of urban settlements}

\author{Erneson A. Oliveira$^{1,2,3}\footnote{Correspondence to:
erneson@fisica.ufc.br}$, Vasco Furtado$^1$, Jos\'e S. Andrade Jr.$^{3}$,
Hern\'an A. Makse$^{3,4}$}

\affiliation{$^1$ Programa de P\'os Gradua\c c\~ao em Inform\'atica Aplicada,
Universidade de Fortaleza, 60811-905 Fortaleza, Cear\'a, Brasil.\\ $^2$ Mestrado
Profissional em Ci\^encias da Cidade, Universidade de Fortaleza, 60811-905
Fortaleza, Cear\'a, Brasil.\\ $^3$ Departamento de F\'isica, Universidade
Federal do Cear\'a, Campus do Pici, 60451-970 Fortaleza, Cear\'a, Brasil.\\ $^4$
Levich Institute and Physics Department, City College of New York, New York, New
York 10031, USA.}

\date{\today}

\begin{abstract}
{\bf The shape of urban settlements plays a fundamental role in their
sustainable planning. Properly defining the boundaries of cities is challenging
and remains an open problem in the Science of Cities. Here, we propose a
worldwide model to define urban settlements beyond their administrative
boundaries through a bottom-up approach that takes into account geographical
biases intrinsically associated with most societies around the world, and
reflected in their different regional growing dynamics. The generality of the
model allows to study the scaling laws of cities at all geographical levels:
countries, continents, and the entire world. Our definition of cities is robust
and holds to one of the most famous results in Social Sciences: Zipf's law.
According to our results, the largest cities in the world are not in line with
what was recently reported by the United Nations. For example, we find that the
largest city in the world is an agglomeration of several small settlements close
to each other, connecting three large settlements: Alexandria, Cairo, and Luxor.
Our definition of cities opens the doors to the study of the economy of cities
in a systematic way independently of arbitrary definitions that employ
administrative boundaries.}
\end{abstract}
\maketitle

\section{Introduction}

What are cities? In {\it The Death and Life of the Great American Cities},
Jacobs argues that human relations can be seen as a proxy for places within
cities \cite{jacobs1961}. A modern view of cities establishes that they can be
defined by the interactions among several types of networks \cite{batty2013,
west2017}, from infrastructure networks to social networks. In recent years, an
increasing number of studies have been proposed to define cities through
consistent mathematical models \cite{makse1995, makse1998, rozenfeld2008,
rozenfeld2011, rybski2013, murcio2013, hernando2014, frasco2014, arcaute2014,
masucci2015, arcaute2016, fluschnik2016} and to investigate urban indicators at
inter- and intra-city scales, in order to shed some light on problems faced by
decision makers \cite{bettencourt2007, arbesman2009, bettencourt2010a,
bettencourt2010b, gomez-lievano2012, gallos2012a, bettencourt2013, fragkias2013,
oliveira2014, melo2014, louf2014, alves2015, li2015, hanley2016, caminha2017,
operti2017}. Despite the efforts of such studies, to properly defining the
boundaries of urban settlements remains an open problem in the Science of
Cities. A minimum criterion of acceptability for any model of cities seems to be
the one that retrieves a conspicuous scaling law found for United States (US),
United Kingdom (UK), and other countries, known as Zipf's law
\cite{rozenfeld2008, rozenfeld2011, gabaix1999a, gabaix1999b, ioannides2003,
cordoba2008, giesen2010, giesen2011, jiang2011, giesen2012, ioannides2013,
duraton2014, watanabe2015}.

In 1949, Zipf \cite{zipf1949} observed that the frequency of words used in the
English language obeys a natural and robust power law behavior, {\it i.e.} a few
words are used many times, while many words are used just a few times. Zipf's
law can be represented generically by the following relation between the size
$S$ of objects from a given set and its rank $R$:
\begin{equation}\label{eq1}
R \propto S^{-\zeta},
\end{equation}

\noindent
where $\zeta=1$ is Zipf's exponent. The size of objects is, in the original
context, the frequency of used words. On the other hand, if such objects are
cities, then the sizes stand for the population of each city, taking into
account Zipf's law and reflecting the fact that there are more small towns than
metropolises in the world. We emphasize that it is not straightforward that the
Zipf's law, despite its robustness, should hold independently of the city
definition, since other scaling relations are not, such as the allometric
exponents for CO$_2$ emissions and light pollution \cite{oliveira2014,
operti2017}. Many other man-made and natural phenomena also exhibit the same
persistent result, {\it e.g.} earthquakes and incomes \cite{sornette1996,
okuyama1999}.

Here, we propose a worldwide model to define urban settlements beyond their
usual administrative boundaries through a bottom-up approach that takes into
account cultural, political, and geographical biases naturally embedded in the
population distribution of continental areas. After all, it is not surprising
that two regions, {\it e.g.} one in western Europe and another one in eastern
Asia, spatially contiguous in population or in commuting level have different
cultural, political or geographical characteristics. Thus, it is also not
surprising that such issues yield different stages of the same mechanics of
growth. The main goal of our model is to be successful in defining cities even
in large regions. Our conjecture is straightforward: there are hierarchical
mechanisms, similar to those present in previous studies of cities in the UK
\cite{arcaute2016} and brain networks \cite{gallos2012b}, behind the growth and
innovation of urban settlements. These mechanisms are ruled by a combination of
general measures, such as the population and the area of each city, and
intrinsic factors which are specific to each region, {\it e.g.} topographical
heterogeneity, political and economic issues, and cultural customs and
traditions. In other words, if political turmoil or economic recession plagues a
metropolis for a long time, all of its satellites are affected too, {\it i.e.}
the entire region ruled by the metropolis will be negatively impacted.

\section{The models}

\subsection{City Clustering Algorithm (CCA)}

In 2008, Rozenfeld {\it et al.} \cite{rozenfeld2008} proposed a model to define
cities beyond their usual administrative boundaries using a notion of spatial
continuity of urban settlements, called the City Clustering Algorithm (CCA)
\cite{rozenfeld2008, rozenfeld2011, rybski2013, oliveira2014, frasco2014,
fluschnik2016, caminha2017, operti2017}. The CCA is defined for discrete or
continuous landscapes \cite{rozenfeld2011} by two parameters: a population
density threshold $D^*$ and a distance threshold $\ell$. These parameters
describe the populated areas and the commuting distance between areas,
respectively. Here, we adopt the following strategy to improve the discrete CCA
performance: (i) Supposing a regular rectangular lattice $L_x \times L_y$ of
sites where the population density of the $k$-th site is $D_k$, we perform an
initial agglomeration by $D^*$ to identify all clusters. If $D_k>D^*$, then the
$k$-th site is populated and we aggregate it with its populated nearest
neighbors. Otherwise, the $k$-th site is unpopulated. (ii) For each populated
cluster, we define its {\it shell sites}, {\it i.e.} sites in the interface
between populated and unpopulated areas. (iii) Lastly, we perform a final
agglomeration by $\ell$, taking into account only the shell sites. If
$d_{ij}<\ell$, where $d_{ij}$ is the distance between the $i$-th and $j$-th
shell sites, and if they belong to different clusters, then the $i$-th and
$j$-th sites belong to the same CCA cluster, even with spatial discontinuity.
Otherwise, they indeed belong to different CCA clusters. This simple strategy
improves the algorithm's computational performance because the number of shell
sites is proportional to $L$, where $L=L_x \approx L_y$ is a linear measure of
the lattice.

\subsection{City Local Clustering Algorithm (CLCA)}

We propose a worldwide model based on the CCA, called the City Local Clustering
Algorithm (CLCA), not only to define cities beyond their usual administrative
boundaries, but also to take into account the intrinsic cultural, political and
geographical biases associated with most societies and reflected in their
particular growing dynamics. The traditional CCA, with fixed $\ell$ and $D^*$,
when applied to a large population density map, can introduce biases defining a
lot of clusters in some regions, while in others just a few. We present the CLCA
with the aim of defining cities even in large regions in order to overcome such
CCA weakness. Hence, it is possible that other models, such as the models based
on street networks proposed by Masucci {\it et al.} \cite{masucci2015} and
Arcaute {\it et al.} \cite{arcaute2016}, carry the same CCA burden and that
local adaptations are necessary for their applications into large regions.

\begin{figure}[htb]
\includegraphics*[width=1.0\textwidth]{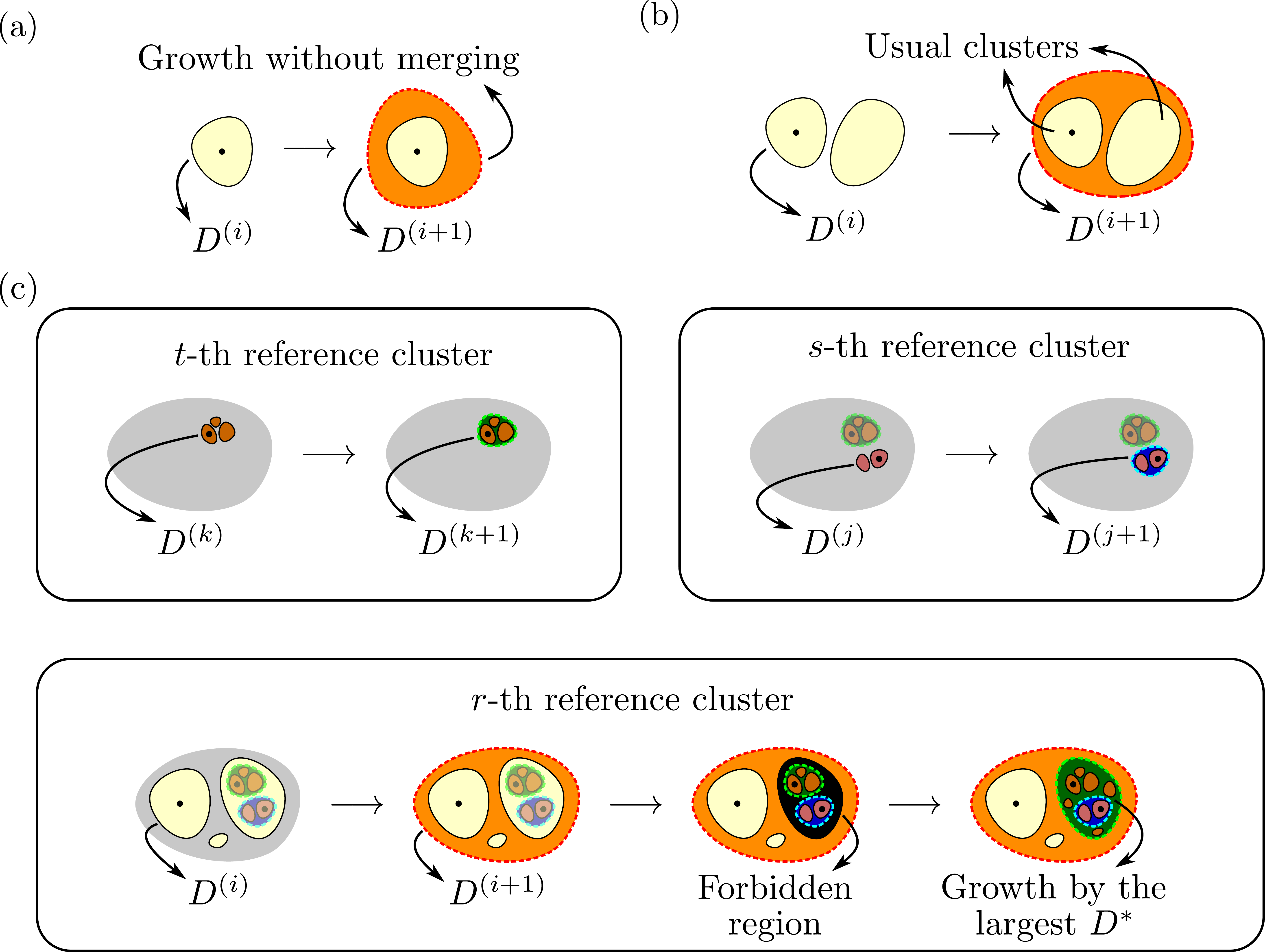}
\caption{{\bf (Color online) City Local Clustering Algorithm (CLCA):
Representation of the conditions (ii) and (iii)}. (a) The growth of the
reference cluster without the merging process. (b) The rising of the usual
clusters. The usual clusters are the CCA clusters at the imminence of the
merging process between $D^{(i)}$ and $D^{(i+1)}$. (c) For $t$-th, $s$-th, and
$r$-th reference clusters ($t$-th is prior to $s$-th which is prior to $r$-th),
the merging processes are performed as described in (b), even though there are
clusters already defined close to and within the current analysed region in the
second and third case, respectively. In the latter, there is the emergence of a
forbidden region. The forbidden regions are the complementary areas of the
reference clusters already defined within the usual clusters. In order to define
the clusters inside those areas, we force the region defined by the largest
value of $D^*$ to grow to the limits of the forbidden region. Here, we suppose
that $D^{(j)}>D^{(k)}$. The filled dots stand for the reference sites.}
\label{clca}
\end{figure}

The main idea of ​​our model is to analyze the change of the CCA clusters
through the variation of $D^*$ under the perspective of different regions.
First, we define a regular rectangular lattice $L_x \times L_y$ of sites, where
the population density of the $k$-th site is $D_k$. We sort all the sites in a
list according to the population density, in descending order. Therefore, the
site with the greatest population density is the first entry in this list, which
we call the first {\it reference site}. The reference site can be considered as
the current core of the analyzed region. Second, we apply the CCA to the
lattice, keeping a fixed value of $\ell$, for a range of $D^*$ decreasing from a
maximum value $D^{(max)}$ to a minimum value $D^{(min)}$ with a decrement
$\delta$. During the decreasing of $D^*$, clusters are formed and they spread
out to all regions of the lattice. Eventually, the cluster that contains the
reference site (from now on the {\it reference cluster}), together with one or
more of the other clusters, will merge from $D^{(i)}$ to $D^{(i+1)}$, where
$D^{(i+1)} = D^{(i)}-\delta$. In order to accept or deny the merging of these
clusters, we introduce three conditions:

\begin{enumerate}[(i)]
\item If the area $A_r(D^{(i)})$ of the reference cluster $r$, {\it i.e.} the
cluster that contains the $r$-th reference site at $D^{(i)}$, obeys
\begin{equation}\label{eq2}
A_r(D^{(i)}) < A^*,
\end{equation}

\noindent
then the reference cluster $r$ always merges with other clusters, because it is
still considered very small. In this context, the area $A^*$ can be understood
as the minimal area of a metropolis.

\item If the difference between the areas of the reference cluster $r$ at
$D^{(i+1)}$ and $D^{(i)}$ obeys
\begin{equation}\label{eq3}
A_r(D^{(i+1)})-A_r(D^{(i)}) > H^* A_r(D^{(i)}),
\end{equation}

\noindent
then the reference cluster $r$ has grown without merging (Fig.\ref{clca}a) or
there is a merging of at least two large clusters (Fig.\ref{clca}b). In the last
case, we emphasize that if there are more than two clusters involved in the
merging process, the reference cluster $r$ may not be one of the largest. As the
first case is not desirable, we can avoid it by reducing the value of $\delta$
and keeping the value of $H^*$ relatively high. The parameter $H^*$ can be
understood as the percentage of the area of the reference cluster $r$ at
$D^{(i)}$. If the second case happens, we consider the entire region inside of
the reference cluster $r$ at $D^{(i+1)}$, but the clusters of this region (which
we call the {\it usual clusters}) are defined by those at $D^{(i)}$. The usual
clusters are the CCA clusters at the imminence of the merging process between
$D^{(i)}$ and $D^{(i+1)}$. This includes the reference cluster $r$ itself and
one or more of the other clusters before the merging (Fig.\ref{clca}b).
Furthermore, all of the sites of the reference cluster $r$ at $D^{(i+1)}$ are
removed from the initial list of reference sites. This condition is necessary
because we should not merge two large metropolises.

\item In condition (ii), when a reference cluster $r$ is merging with another
cluster that covers one or more regions already defined by previous reference
clusters at different values of $D^*$, there is a strong likelihood of the
emergence of a {\it forbidden region} within that cluster. In this case, we
force the region already defined by the largest value of $D^*$ to grow to the
limits of the forbidden region (Fig.\ref{clca}c). The forbidden regions are the
complementary areas of the reference clusters already defined within the usual
clusters. As a consequence of this procedure, some CCA clusters that were hidden
after the analysis of the previous reference cluster arise in this forbidden
region. We justify this condition by the idea that a metropolis rules the growth
of its satellites, since it plays a fundamental role in their socioeconomic
relations.
\end{enumerate}

\noindent
We apply the same procedure to the second reference cluster, to the third
reference cluster, and so on. Finally, we also define the {\it isolated
clusters} with the minimum value of $D^*$ for all the cases accepted in
condition (ii). In order to make our model clearer, we chose the descending
order to sort the population density for one reason: To favor the merging
process of the high density clusters that rose from the decreasing of the $D^*$.
In practice, we run our revised discrete CCA just once for the entire range of
input parameters and store all of the outputs in order to improve the
performance of the model. The apparent simplicity of this task hides a RAM
memory management problem of storing all of the outputs in a medium-performance
computer. We overcome such a barrier through the {\it zram module}
\cite{zram2017}, available in the newest linux kernels. The zram module creates
blocks which compress and store information dynamically in the RAM memory
itself, at the cost of processing time.

\section{The dataset}

\begin{figure}[htb]
\includegraphics*[width=0.8\textwidth]{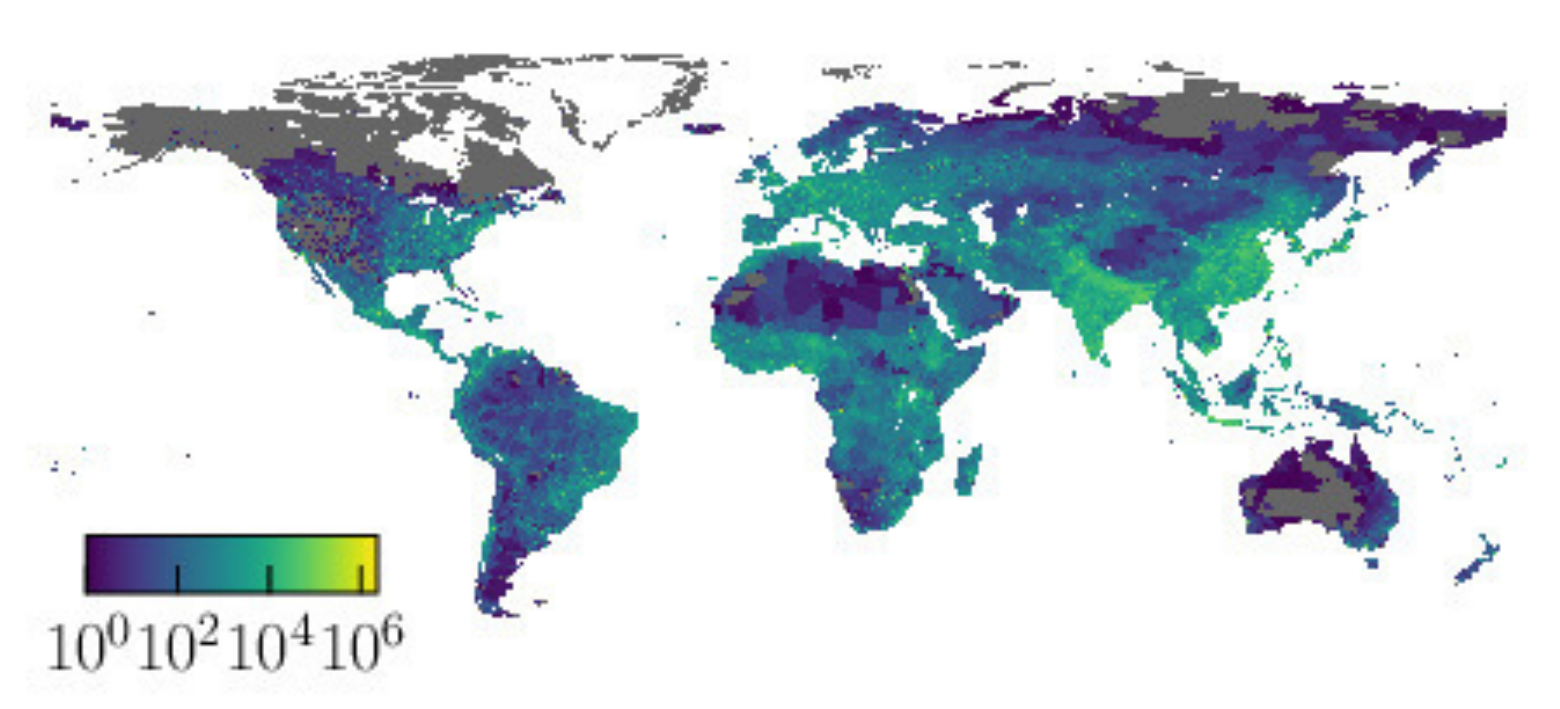}
\caption{{\bf (Color online) The Global Rural-Urban Mapping Project (GRUMPv1)
dataset.} The population map of the entire world from the GRUMPv1 dataset in
logarithmic scale.}
\label{grumpv1}
\end{figure}

We use the {\it Global Rural-Urban Mapping Project} (GRUMPv1) \cite{grump2011},
available from the {\it Socioeconomic Data and Applications Center} (SEDAC) at
Columbia University, to apply the CLCA to a single global dataset. The GRUMPv1
dataset is composed of georeferenced rectangular population grids for 232
countries around the world in the year 2000 (Fig. \ref{grumpv1}). Such a dataset
is a compilation of gridded census and satellite data for the populations of
urban and rural areas. These data are provided at a high resolution of
$30\;\text{arc-seconds}$, equivalent to $30/3600\;\text{degrees}$ or a grid of
$0.926\;km \times 0.926\;km$ at the Equator. We note that despite of the
heterogeneous population distributions that built the GRUMPv1, its overall
resolution is tolerable to the CLCA, since we can identify well-defined clusters
around all continents in the raw data.

We calculate the area of ​​each site by the composition of two spherical
triangles \cite{snyder1987}. The area of a spherical triangle with edges $a$,
$b$ and $c$ is given by
\begin{equation}\label{eq4}
A=4 R_e^2 \tan^{-1}\left[\tan\left(\frac{s}{2}\right)\tan\left(\frac{s_a}{2}\right)\tan\left(\frac{s_b}{2}\right)\tan\left(\frac{s_c}{2}\right)\right]^{1/2},
\end{equation}

\noindent
where $s=(a/R_e+b/R_e+c/R_e)/2$, $s_a=s-a/R_e$, $s_b=s-b/R_e$, and
$s_c=s-c/R_e$. In this formalism, $R_e=6,378.137\;km$ is the Earth's radius and
the edge lengths are calculated by the great circle (geodesic) distance between
two points $i$ and $j$ on the Earth's surface:
\begin{equation}\label{eq5}
d_{ij}=R_e\cos^{-1}[\sin(\phi_i) \sin(\phi_j)+\cos(\phi_i)\cos(\phi_j)\cos(\lambda_j-\lambda_i)].
\end{equation}

\noindent
The values of $\lambda_i$ ($\lambda_j$) and $\phi_i$ ($\phi_j)$, measured in
radians, are the longitude and latitude, respectively, of the point $i$ ($j$).
Thus, we are able to define the population density for each site of the lattice,
since its population and area are known.

\begin{figure}[htb]
\includegraphics*[width=0.8\textwidth]{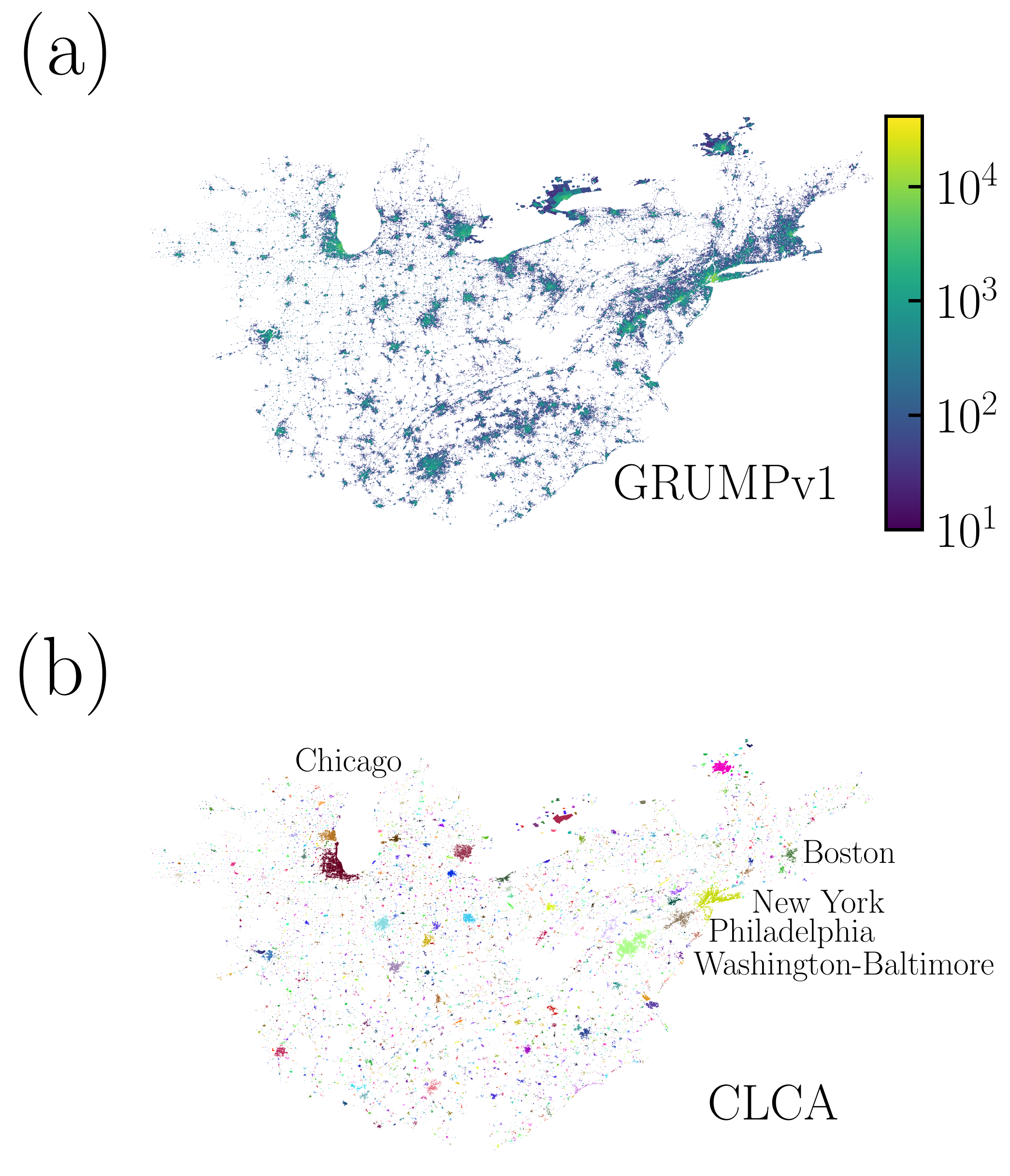}
\caption{{\bf (Color online) The largest cluster of regions for the United
States (US).} (a) The single population density cluster from the Eastern US is
defined by the City Clustering Algorithm (CCA) with lower and upper bound
parameters $D^*=50\;\text{people}/km^2$ and $\ell=10\;km$, respectively. The
population, provided by the Global Rural-Urban Mapping Project (GRUMPv1)
dataset, is shown in logarithmic scale within each populated area. (b)
Application of the City Local Clustering Algorithm (CLCA) for the cluster of
regions of the Eastern US. The CLCA cities are represented in several colors,
{\it e.g.} New York in mustard, Philadelphia in light brown,
Washington-Baltimore in light green, Boston in green and Chicago in red. The
CLCA parameters used were $D^{(min)}=100\;\text{people}/km^2$,
$D^{(max)}=1000\;\text{people}/km^2$, $\delta=10\;\text{people}/km^2$,
$\ell=3\;km$, $A^*=50\;km^2$, and $H^*=0.05$.}
\label{us_east}
\end{figure}

We also pre-process the GRUMPv1 dataset, dividing all countries and continents
--- and even the entire world --- into large regions which we call {\it clusters
of regions}, to apply our model in a feasible computational time using
medium-performance computers. These regions are defined by the CCA with lower
and upper bound parameters $D^*=50\;\text{people}/km^2$ and $\ell=10\;km$,
respectively. We believe that such large clusters can hold the socioeconomic and
cultural relations among different urban settlements of a territory. Fig.
\ref{us_east}a shows the largest clusters of regions in the US; as we can see,
all of the Eastern US is considered a single cluster.

\section{Results}

To show the relevance of our model, we apply the CLCA to the GRUMPv1 dataset at
three different geographic levels: countries, continents and the entire world.
For each case, we consider only a single set of CLCA parameters. We justify our
choices with the following assumptions: (i) $D^{(min)}=100\;\text{people}/km^2$,
a value slightly greater than the lower bound CCA parameter
($D^*=50\;\text{people}/km^2$) used to define the regions of clusters; (ii)
$D^{(max)}=1000\;\text{people}/km^2$, a loosened value of $D^{(max)}=\infty$;
(iii) $\delta=10\;\text{people}/km^2$, a small enough value to avoid the
reference clusters growing without merging; (iv) $\ell=3\;km$, the critical
distance threshold, already extensively analyzed by previous CCA studies
\cite{rozenfeld2008, rozenfeld2011, oliveira2014}; (v) $A^*=50\;km^2$, the
minimum area of a metropolis, as it is required that $A^*$ be reasonably greater
than the minimum unit of area from the dataset and smaller than a metropolis'
area; and (vi) $H^*=0.05$, a large enough value to favor the merging of clusters
which are similar in size. The Fig. \ref{us_east}b shows the CLCA cities defined
by the single set of CLCA parameters. For other regions, see the Supplementary
Information (SI).

\begin{figure}[htb]
\includegraphics*[width=0.8\textwidth]{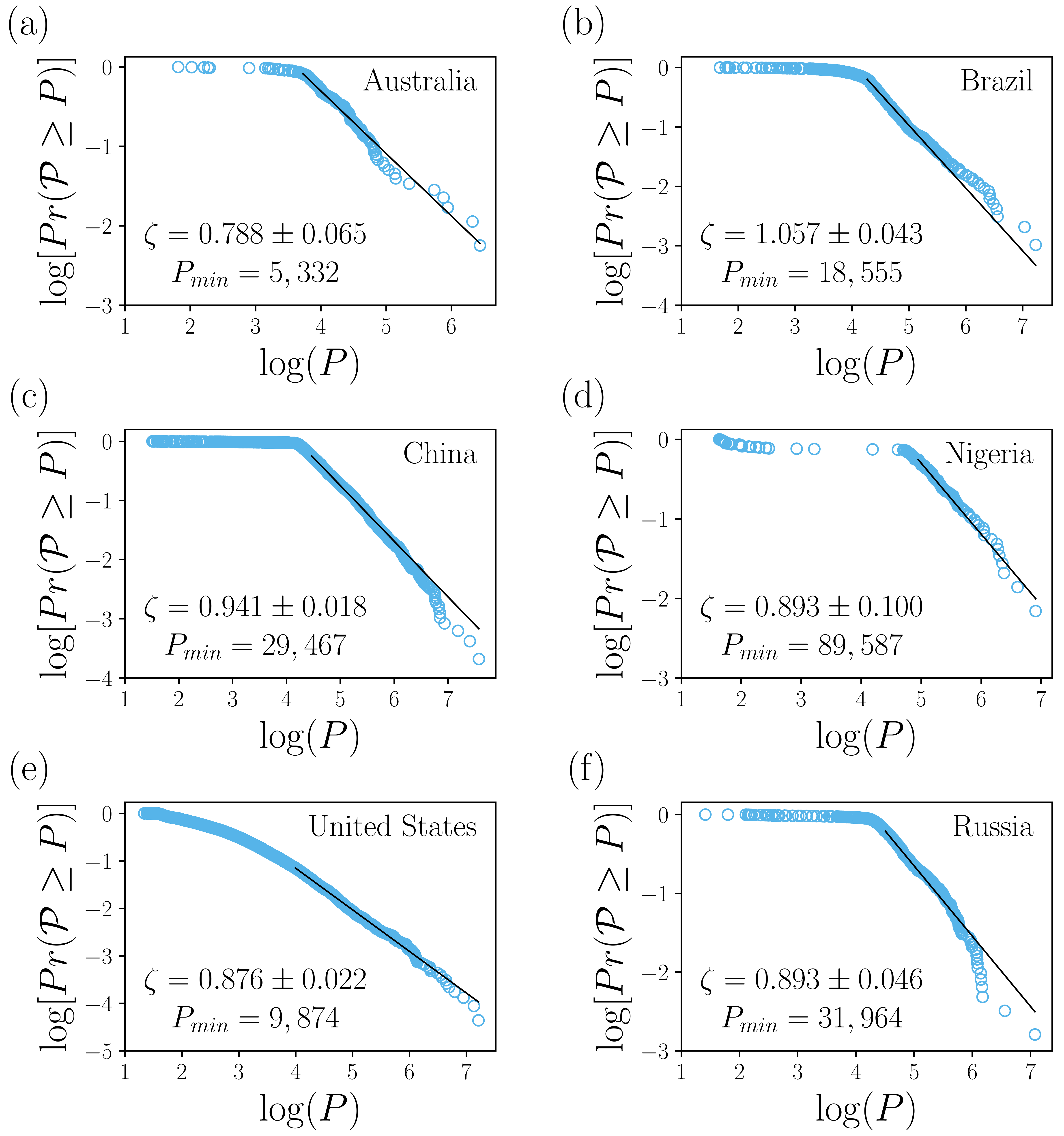}
\caption{{\bf (Color online) Cumulative Distribution Function (CDF)
$Pr(\mathcal{P} \geq P)$ versus population $P$, in log-log scale, for the
countries with the highest number of cities in each continent (for other
countries, see the SI).} (a)-(f) Cities proposed by the City Local Clustering
Algorithm (CLCA) are represented by light blue circles. The solid black line is
the maximum likelihood power-law fit defined by the Maximum Likelihood Estimator
(MLE) \cite{clauset2009}. The value of the lower bound $P_{min}$ and the
exponent $\zeta$ are also shown. The CLCA parameters used were
$D^{(min)}=100\;\text{people}/km^2$, $D^{(max)}=1000\;\text{people}/km^2$,
$\delta=10\;\text{people}/km^2$, $\ell=3\;km$, $A^*=50\;km^2$ and $H^*=0.05$.}
\label{g_countries_panel}
\end{figure}

We study the population distribution using the Maximum Likelihood Estimator
(MLE) proposed by Clauset {\it et al.} \cite{clauset2009}. Their approach
combines maximum-likelihood fitting methods with goodness-of-fit tests based on
Kolmogorov-Smirnov (KS) statistic. The Fig. \ref{g_countries_panel} shows the
log-log behavior of the Cumulative Distribution Function (CDF) for the
population of the CLCA cities, considering only the countries with the highest
number of CLCA cities for each continent (for other countries, see the SI). The
$Pr(\mathcal{P} \geq P)$ represents the probability that a random population
$\mathcal{P}$ takes on a value greater or equal to the population $P$. In all
CDF plots, we also show the maximum likelihood power-law fit, as well as, the
value of the exponent $\zeta=\alpha-1$, where $\alpha$ is the MLE exponent, and
the value of $P_{min}$, the lower bound of the MLE.

\begin{figure}[htb]
\includegraphics*[width=0.8\textwidth]{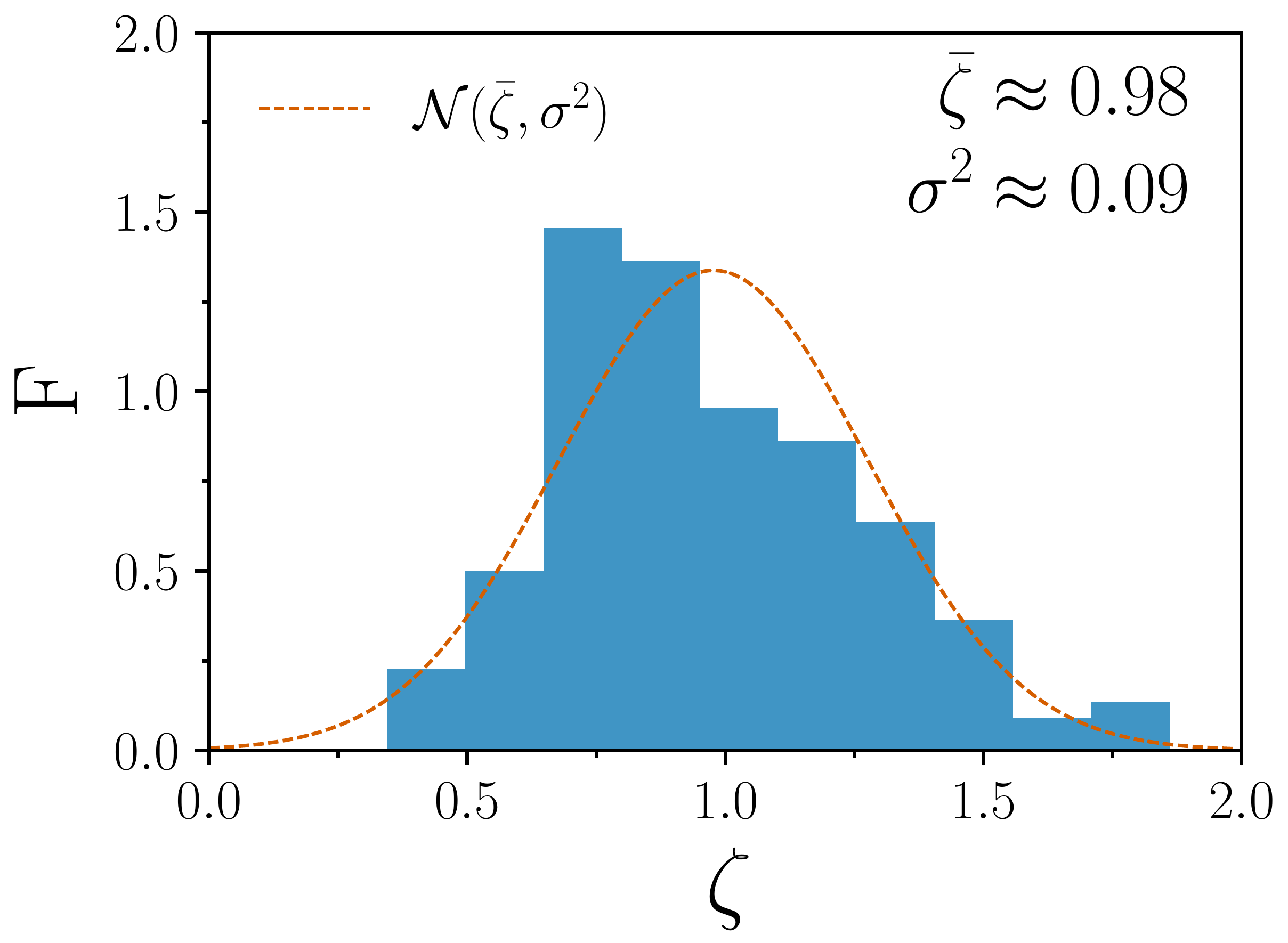}
\caption{{\bf (Color online) Normalized histogram, with frequency $F$, of the
$\zeta$ exponent at the country level.} The plot shows those countries (145 out
of 232) with at least 10 cities defined by the City Local Clustering Algorithm
(CLCA) in the region covered by the maximum likelihood power-law fit. We find
the mean value of the Zipf's exponents $\bar{\zeta}=0.98$ and its variance
$\sigma^2=0.09$. The dashed red line stands for the normal distribution
$\mathcal{N}(\bar{\zeta},\sigma^2)$. Therefore, Zipf's law holds for the most
countries.}
\label{g_countries_hist}
\end{figure}

In Fig. \ref{g_countries_hist}, we show a normalized histogram, with frequency
$F$, of the $\zeta$ exponents for all countries ( 145 out of 232) with at least
10 CLCA cities in the region covered by the maximum likelihood power-law fit.
The mean value of the $\zeta$ exponents is $\bar{\zeta}=0.98$, with variance
$\sigma^2=0.09$. The dashed red line stands for the normal distribution
$\mathcal{N}(\bar{\zeta},\sigma^2)$. In spite of the $\zeta$ exponent
heterogeneity illustrated by Fig. \ref{g_countries_hist}, Zipf's law holds for
most countries around the globe. We emphasize that such results corroborate with
previous studies performed for one country or a small number of countries
\cite{rozenfeld2008, rozenfeld2011, gabaix1999a, gabaix1999b, ioannides2003,
cordoba2008, giesen2010, giesen2011, jiang2011, giesen2012, ioannides2013,
duraton2014, watanabe2015}. In special, the Fig. \ref{g_countries_hist} also
endorses an astute meta-analysis performed by Cottineau \cite{cottineau2017}.
Cottineau provided a comparison among the Zipf's law exponents found in 86
studies. Our results strongly corroborate those presented in such study, except
that our exponents are ranged between 0 and 2.

\begin{figure}[htb]
\includegraphics*[width=0.8\textwidth]{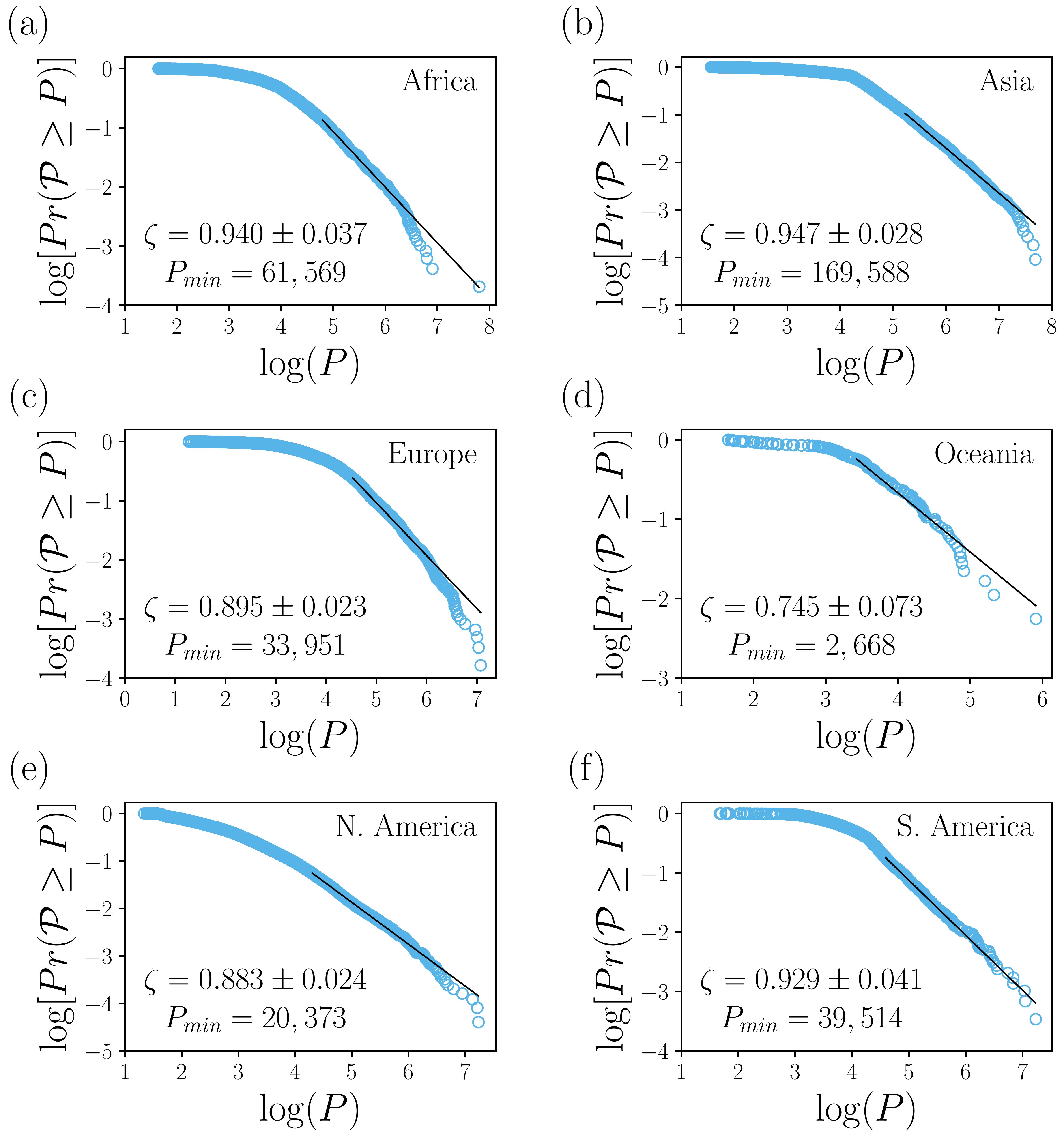}
\caption{{\bf (Color online) Cumulative Distribution Function (CDF)
$Pr(\mathcal{P} \geq P)$ versus population $P$, in log-log scale, for the
continents.} (a)-(f) Cities proposed by the City Local Clustering Algorithm
(CLCA) are represented by light blue circles. The solid black line is the
maximum likelihood power-law fit defined by the Maximum Likelihood Estimator
(MLE) \cite{clauset2009}. The value of the lower bound $P_{min}$ and the
exponent $\zeta$ are also shown. The CLCA parameters used were
$D^{(min)}=100\;\text{people}/km^2$, $D^{(max)}=1000\;\text{people}/km^2$,
$\delta=10\;\text{people}/km^2$, $\ell=3\;km$, $A^*=50\;km^2$ and $H^*=0.05$.}
\label{g_continents_panel}
\end{figure}

\begin{figure}[htb]
\includegraphics*[width=0.8\textwidth]{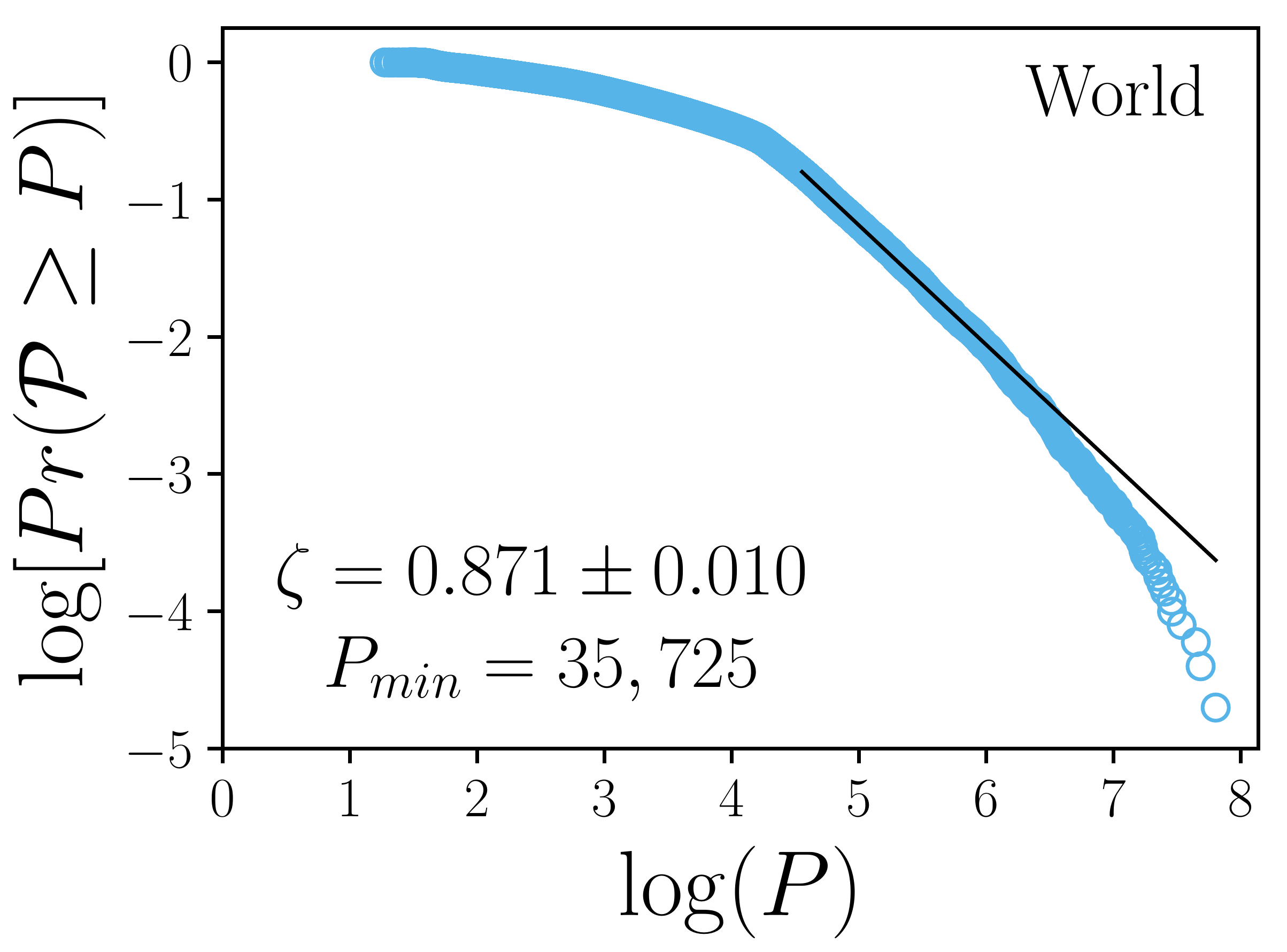}
\caption{{\bf (Color online) Cumulative Distribution Function (CDF)
$Pr(\mathcal{P} \geq P)$ versus population $P$, in log-log scale, for the entire
world.} (a)-(f) Cities proposed by the City Local Clustering Algorithm (CLCA)
are represented by light blue circles. The solid black line is the maximum
likelihood power-law fit defined by the Maximum Likelihood Estimator (MLE)
\cite{clauset2009}. The value of the lower bound $P_{min}$ and the exponent
$\zeta$ are also shown. The CLCA parameters used were
$D^{(min)}=100\;\text{people}/km^2$, $D^{(max)}=1000\;\text{people}/km^2$,
$\delta=10\;\text{people}/km^2$, $\ell=3\;km$, $A^*=50\;km^2$ and $H^*=0.05$.}
\label{g_world}
\end{figure}

Furthermore, we challenge the robustness of our model at higher geographic
levels: continents and the entire world. We performed the same analyses and find
that our results persist on both scales, {\it i.e.} the CLCA cities follows
Zipf's law for continents and the entire world, as illustrated in Figs.
\ref{g_continents_panel} and \ref{g_world}.

We summarize our results in a set of 7 tables: Tables \ref{t_countries_af},
\ref{t_countries_as}, \ref{t_countries_eu}, \ref{t_countries_na} ,
\ref{t_countries_oc}, and \ref{t_countries_sa}, for countries from Africa; Asia;
Europe; North America; Oceania; and South America, respectively. Table
\ref{t_continents_world} contains similar information for all continents and the
entire world. In all cases, we show the name of the considered region (country,
continent or globe), the ISO 3166-1 alpha-3 code associated (only for
countries), the number of cities obtained by the CLCA and those covered by the
MLE, the lower bound $P_{min}$, and the Zipf's exponent $\zeta$.

\begin{figure}[htb]
\includegraphics*[width=0.8\textwidth]{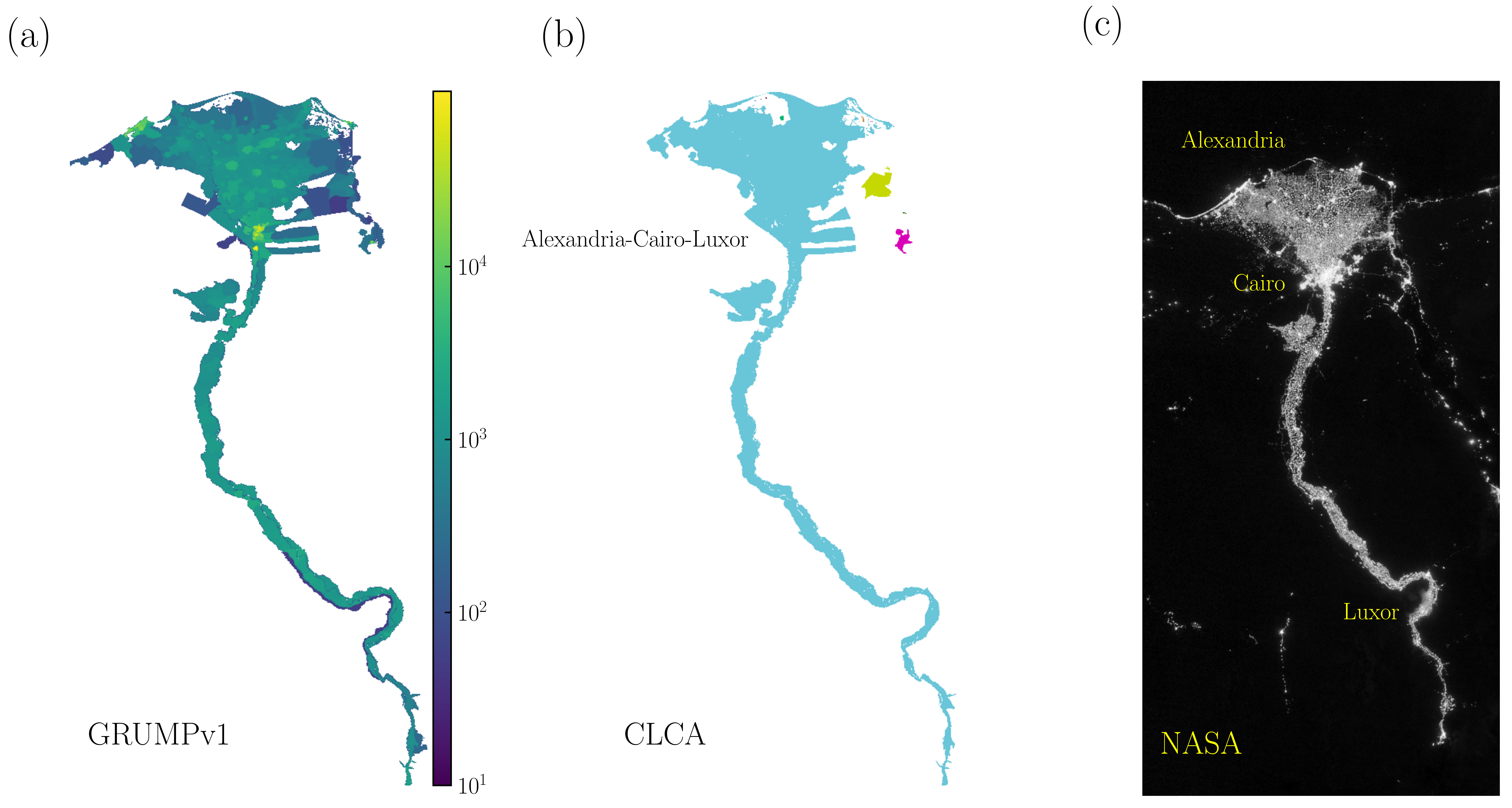}
\caption{{\bf (Color online) Northeastern region of Egypt}. (a) The cluster of
regions defined by the pre-processing of the GRUMPv1 dataset for the
Northeastern region of Egypt. (b) The largest city defined by the City Local
Clustering Algorithm (CLCA) in the entire world is formed by several cities,
including Alexandria, Cairo, and Luxor. (c) Night-time lights of the Northeast
of Egypt provided by National Aeronautics and Space Administration (NASA). The
CLCA cities found exhibit a remarkable similarity with the lights across the
Nile.}
\label{cairo}
\end{figure}

It is remarkable that the top CLCA city, with a population of 63,585,039 people,
is composed of three large urban settlements (Alexandria, Cairo, and Luxor)
connected by several small ones. Figs. \ref{cairo}a, \ref{cairo}b, and
\ref{cairo}c show the largest cluster of regions in Egypt for the GRUMPv1
dataset; CLCA cities; and night-time lights from the National Aeronautics and
Space Administration (NASA) \cite{nasa2016}, respectively. We believe the main
reason for this finding has been present in the Northeast of Africa since before
the beginning of ancient civilization --- namely, the Nile river. Actually, it
is well known that almost the entire Egypt population lives in a strip along the
Nile river, in the Nile delta, and in the Suez canal on 4\% of the total country
area ($~10^6\;km^2$), where there are arable lands to produce food
\cite{gehad2003}. The river and delta regions are composed by some large cities
and a lot of small villages, making them extremely dense. Therefore, our results
raise the hypothesis that the cities and villages across the Nile can be seen as
a kind of ``megacity'', despite spatially non-contiguous, due to the
socioeconomic relation, reflected on the high commuting levels, among close
subregions.

The Table \ref{t_world} shows the top 25 CLCA cities in the entire world by
population, and their associated areas. After the top CLCA city,
Alexandria-Cairo-Luxor, we emphasize that the 13 next-largest CLCA cities are in
Asia. Indeed, we can see that the shape of the tail end of the entire world
population distribution (in Fig. \ref{g_world}) is roughly ruled by the greater
CLCA city in Africa and several CLCA cities in Asia. 

These facts are not in line with what was recently reported by the {\bf United Nations (UN)} \cite{un2016}, {\it e.g.} the largest CLCA city, Alexandria-Cairo-Luxor, is just the 9th largest city according to the UN, and the largest UN city, Tokyo, is just the 4th largest according to our analyses.

\section{Conclusions}

We propose a model to define urban settlements through a bottom-up approach
beyond their usual administrative boundaries, and moreover to account for the
intrinsic cultural, political, and geographical biases associated with most
societies and reflected in their particular growing dynamics. We claim that such
a property qualifies our model to be applied worldwide, without any regional
restrictions. We also propose an alternative strategy to improve the
computational performance of the discrete CCA. We emphasize that the CCA can
still be used to define cities; however, it depends upon a different tuning of
its parameters for each large region without direct socioeconomic and political
relations. Furthermore, we show that the definition of cities proposed by our
approach is robust and holds to one of the most famous results in Social
Science, Zipf's law, not only for previously studied countries, {\it e.g.} the
US, the UK, or China, but for all countries (145 from 232 provided by GRUMPv1)
around the world. We also find that Zipf's law emerges at different geographic
levels, such as continents and the entire world. Another highlight of our study
is the fact that our model is applied upon one single dataset to define all
cities. Furthermore, we find that the most populated cities are not the major
players in the global economy (such as New York City, London, or Tokyo). The
largest CLCA city, with a population of 63,585,039 people, is an agglomeration
of several small cities close to each other which connects three large cities:
Alexandria, Cairo, and Luxor. Finally, after the top CLCA city of
Alexandria-Cairo-Luxor, we find that the next-largest 13 CLCA cities are in
Asia. These facts are not in full agreement with a recent UN report
\cite{un2016}. According to our results, the largest CLCA city,
Alexandria-Cairo-Luxor, is just the 9th largest city according to the UN, while
the largest UN city, Tokyo, is just the 4th largest according to our analyses.


\section{Data Accessibility}
The dataset supporting this article are available at
\url{http://sedac.ciesin.columbia.edu/data/collection/grump-v1}. More
specifically, the reader can click on ``Data sets'' and, after that, on
``Population Count Grid, v1 (1990,1995,2000)''. We also provide the codes for
the proposed model that are available at
\url{https://doi.org/10.5061/dryad.968nq8n}.

\section{Competing Interests}
We have no competing interests.

\section{Author's Contributions}
EAO performed the data analysis, the algorithm of the proposed model, and the
statistical analysis. He also participated in the design of the study and
drafted the manuscript. VF carried out the funding acquisition and helped draft
the manuscript. JSA participated in the design of the study, carried out the
funding acquisition, and helped draft the manuscript. HAM conceived, designed,
and coordinated the research, as well as, carried out the funding acquisition
and helped draft the manuscript. All authors approved the manuscript.

\section{Acknowledgements}
We thank the Global Rural-Urban Mapping Project (GRUMPv1) team for the dataset
provided. Furthermore, we would like to thank X. Gabaix for helpful comments and
discussions.

\section{Funding}
We gratefully acknowledge funding by CNPq, CAPES, FUNCAP, NSF, and ARL
Cooperative Agreement Number W911NF-09-2-0053 (the ARL Network Science CTA).


\clearpage

\clearpage

\clearpage

\clearpage

\clearpage

\clearpage

\clearpage

\clearpage

\clearpage

\LTcapwidth=\textwidth
\begin{longtable}{c|c|c|c|c|c}
Country & ISO & CLCA cities & CLCA cities$^\dagger$ & $P_{min}$ & $\zeta$\\ \hline
Angola & AGO & 20 & 16 & 43,937 & 0.780 $\pm$ 0.195\\
Benin & BEN & 40 & 30 & 12,607 & 0.780 $\pm$ 0.142\\ 
Burkina Faso & BFA & 139 & 78 & 12,314 & 1.256 $\pm$ 0.142\\ 
Botswana & BWA & 79 & 58 & 1,674 & 0.785 $\pm$ 0.103\\ 
Central African Republic & CAF & 37 & 11 & 14,868 & 1.230 $\pm$ 0.371\\ 
Ivory Coast & CIV & 83 & 47 & 18,400 & 0.962 $\pm$ 0.140\\ 
Cameroon & CMR & 143 & 93 & 7,478 & 0.711 $\pm$ 0.074\\ 
Democratic Republic of the Congo & COD & 191 & 47 & 25,996 & 0.764 $\pm$ 0.111\\ 
Congo & COG & 21 & 18 & 17,673 & 1.050 $\pm$ 0.248\\ 
Comoros & COM & 16 & 15 & 4,167 & 0.922 $\pm$ 0.238\\ 
Cape Verde & CPV & 16 & 11 & 5,205 & 1.083 $\pm$ 0.327\\ 
Algeria & DZA & 273 & 112 & 24,192 & 0.910 $\pm$ 0.086\\ 
Egypt & EGY & 19 & 12 & 11,967 & 0.511 $\pm$ 0.147\\ 
Eritrea & ERI & 27 & 12 & 6,559 & 0.730 $\pm$ 0.211\\ 
Ethiopia & ETH & 244 & 147 & 6,638 & 0.688 $\pm$ 0.057\\ 
Gabon & GAB & 33 & 27 & 3,108 & 0.844 $\pm$ 0.162\\ 
Ghana & GHA & 95 & 25 & 54,662 & 1.145 $\pm$ 0.229\\ 
Guinea & GIN & 34 & 13 & 40,118 & 1.234 $\pm$ 0.342\\ 
Gambia & GMB & 35 & 33 & 1,186 & 0.610 $\pm$ 0.106\\ 
Guinea-Bissau & GNB & 26 & 14 & 9,148 & 1.139 $\pm$ 0.305\\ 
Kenya & KEN & 179 & 20 & 72,756 & 1.383 $\pm$ 0.309\\ 
Liberia & LBR & 42 & 19 & 6,468 & 0.604 $\pm$ 0.139\\ 
Libyan Arab Jamahiriya & LBY & 30 & 18 & 40,273 & 1.180 $\pm$ 0.278\\ 
Lesotho & LSO & 14 & 11 & 1,999 & 0.651 $\pm$ 0.196\\ 
Morocco (includes Western Sahara) & MAR & 58 & 50 & 26,325 & 0.763 $\pm$ 0.108\\ 
Madagascar & MDG & 138 & 74 & 14,867 & 1.340 $\pm$ 0.156\\ 
Mali & MLI & 152 & 146 & 4,463 & 1.161 $\pm$ 0.096\\ 
Mozambique & MOZ & 127 & 14 & 128,214 & 1.861 $\pm$ 0.497\\ 
Malawi & MWI & 179 & 72 & 4,194 & 0.779 $\pm$ 0.092\\ 
Namibia & NAM & 31 & 17 & 12,467 & 1.637 $\pm$ 0.397\\ 
Niger & NER & 58 & 36 & 10,717 & 0.753 $\pm$ 0.126\\ 
Nigeria & NGA & 144 & 80 & 89,587 & 0.893 $\pm$ 0.100\\ 
Sudan & SDN & 77 & 56 & 39,764 & 1.031 $\pm$ 0.138\\ 
Senegal & SEN & 42 & 34 & 13,475 & 0.798 $\pm$ 0.137\\ 
Sierra Leone & SLE & 62 & 52 & 1,899 & 0.612 $\pm$ 0.085\\ 
Chad & TCD & 75 & 14 & 19,574 & 1.086 $\pm$ 0.290\\ 
Togo & TGO & 54 & 11 & 82,964 & 1.667 $\pm$ 0.503\\ 
Tunisia & TUN & 46 & 36 & 16,130 & 1.014 $\pm$ 0.169\\ 
United Republic of Tanzania & TZA & 114 & 33 & 73,621 & 0.936 $\pm$ 0.163\\ 
Uganda & UGA & 155 & 33 & 30,587 & 1.386 $\pm$ 0.241\\ 
South Africa & ZAF & 1,915 & 97 & 53,320 & 1.270 $\pm$ 0.129\\ 
Zambia & ZMB & 55 & 34 & 7,118 & 0.666 $\pm$ 0.114\\ 
Zimbabwe & ZWE & 28 & 24 & 13,411 & 0.746 $\pm$ 0.152\\
\caption{{\bf African countries.} We show the name, the ISO 3166-1 alpha-3 code,
the number of cities obtained by the City Local Clustering Algorithm (CLCA) and
the number of those covered by the maximum likelihood power-law fit defined by
the Maximum Likelihood Estimator (MLE) \cite{clauset2009} (represented by
$\dagger$), the lower bound $P_{min}$, and the Zipf's exponent $\zeta$.}
\label{t_countries_af}\\
\end{longtable}
\clearpage

\LTcapwidth=\textwidth
\begin{longtable}{c|c|c|c|c|c}
Country & ISO & CLCA cities & CLCA cities$^\dagger$ & $P_{min}$ & $\zeta$\\ \hline
Afghanistan & AFG & 95 & 38 & 29,242 & 0.809 $\pm$ 0.131\\ 
Armenia & ARM & 41 & 19 & 17,088 & 1.256 $\pm$ 0.288\\ 
Azerbaijan & AZE & 34 & 21 & 17,169 & 0.776 $\pm$ 0.169\\ 
Bangladesh & BGD & 103 & 58 & 26,586 & 0.581 $\pm$ 0.076\\ 
Bhutan & BTN & 19 & 15 & 893 & 0.469 $\pm$ 0.121\\ 
China & CHN & 4,782 & 2,706 & 29,467 & 0.941 $\pm$ 0.018\\ 
Cyprus & CYP & 17 & 15 & 626 & 0.486 $\pm$ 0.126\\ 
Georgia & GEO & 52 & 38 & 6,526 & 0.765 $\pm$ 0.124\\ 
Indonesia & IDN & 2,416 & 542 & 12,876 & 0.894 $\pm$ 0.038\\ 
India & IND & 1,040 & 299 & 94,976 & 0.786 $\pm$ 0.045\\ 
Iran & IRN & 169 & 56 & 100,763 & 1.194 $\pm$ 0.160\\ 
Israel & ISR & 24 & 20 & 877 & 0.448 $\pm$ 0.100\\ 
Jordan & JOR & 13 & 11 & 15,253 & 0.803 $\pm$ 0.242\\ 
Japan & JPN & 270 & 33 & 289,039 & 1.011 $\pm$ 0.176\\ 
Kazakhstan & KAZ & 77 & 22 & 103,289 & 1.505 $\pm$ 0.321\\ 
Kyrgyz Republic & KGZ & 134 & 37 & 9,117 & 0.991 $\pm$ 0.163\\ 
Cambodia & KHM & 84 & 24 & 34,495 & 1.735 $\pm$ 0.354\\ 
Korea & KOR & 131 & 23 & 126,819 & 0.750 $\pm$ 0.156\\ 
Lao Peoples Democratic Republic & LAO & 35 & 20 & 12,595 & 0.958 $\pm$ 0.214\\ 
Sri Lanka & LKA & 23 & 20 & 8,573 & 0.466 $\pm$ 0.104\\ 
Maldives & MDV & 149 & 40 & 1,498 & 1.799 $\pm$ 0.285\\ 
Myanmar & MMR & 115 & 37 & 69,935 & 1.190 $\pm$ 0.196\\ 
Mongolia & MNG & 24 & 19 & 13,179 & 1.419 $\pm$ 0.325\\ 
Malaysia & MYS & 119 & 15 & 157,843 & 1.286 $\pm$ 0.332\\ 
Nepal & NPL & 39 & 22 & 15,396 & 0.560 $\pm$ 0.119\\ 
Oman & OMN & 28 & 12 & 34,956 & 1.519 $\pm$ 0.438\\ 
Pakistan & PAK & 96 & 45 & 90,356 & 0.790 $\pm$ 0.118\\ 
Philippines & PHL & 352 & 38 & 106,854 & 1.195 $\pm$ 0.194\\ 
Democratic Peoples Republic of Korea & PRK & 53 & 20 & 174,121 & 1.502 $\pm$ 0.336\\ 
Saudi Arabia & SAU & 57 & 15 & 156,672 & 0.861 $\pm$ 0.222\\ 
Syrian Arab Republic & SYR & 39 & 20 & 29,908 & 0.647 $\pm$ 0.145\\ 
Thailand & THA & 100 & 24 & 23,482 & 0.718 $\pm$ 0.147\\ 
Tajikistan & TJK & 39 & 13 & 17,660 & 0.740 $\pm$ 0.205\\ 
Turkmenistan & TKM & 30 & 14 & 26,319 & 0.883 $\pm$ 0.236\\ 
East Timor & TLS & 23 & 15 & 1,220 & 0.547 $\pm$ 0.141\\ 
Turkey & TUR & 338 & 244 & 18,389 & 0.926 $\pm$ 0.059\\ 
Taiwan & TWN & 16 & 13 & 2,186 & 0.344 $\pm$ 0.095\\ 
Uzbekistan & UZB & 56 & 36 & 15,865 & 0.574 $\pm$ 0.096\\ 
Viet Nam & VNM & 345 & 72 & 35,980 & 0.876 $\pm$ 0.103\\ 
Yemen & YEM & 46 & 22 & 38,276 & 1.059 $\pm$ 0.226\\
\caption{{\bf Asian countries.} We show the name, the ISO 3166-1 alpha-3 code,
the number of cities obtained by the City Local Clustering Algorithm (CLCA) and
the number of those covered by the maximum likelihood power-law fit defined by
the Maximum Likelihood Estimator (MLE) \cite{clauset2009} (represented by
$\dagger$), the lower bound $P_{min}$, and the Zipf's exponent $\zeta$.}
\label{t_countries_as}\\
\end{longtable}
\clearpage

\LTcapwidth=\textwidth
\begin{longtable}{c|c|c|c|c|c}
Country & ISO & CLCA cities & CLCA cities$^\dagger$ & $P_{min}$ & $\zeta$\\ \hline
Albania & ALB & 46 & 32 & 6,030 & 0.783 $\pm$ 0.139\\ 
Austria & AUT & 116 & 74 & 4,383 & 0.754 $\pm$ 0.088\\ 
Belgium & BEL & 43 & 31 & 9,800 & 0.706 $\pm$ 0.127\\ 
Bulgaria & BGR & 56 & 29 & 33,338 & 1.308 $\pm$ 0.243\\ 
Bosnia-Herzegovina & BIH & 57 & 17 & 15,708 & 1.186 $\pm$ 0.288\\ 
Belarus & BLR & 36 & 17 & 73,682 & 1.123 $\pm$ 0.272\\ 
Switzerland & CHE & 71 & 15 & 55,878 & 1.167 $\pm$ 0.301\\ 
Czech Republic & CZE & 206 & 33 & 41,254 & 1.393 $\pm$ 0.243\\ 
Germany & DEU & 331 & 242 & 13,926 & 0.811 $\pm$ 0.052\\ 
Denmark & DNK & 134 & 85 & 2,248 & 0.682 $\pm$ 0.074\\ 
Spain & ESP & 358 & 36 & 133,759 & 1.192 $\pm$ 0.199\\ 
Estonia & EST & 51 & 13 & 14,041 & 1.178 $\pm$ 0.327\\ 
Finland & FIN & 72 & 22 & 27,831 & 1.444 $\pm$ 0.308\\ 
France & FRA & 1,253 & 114 & 42,160 & 1.087 $\pm$ 0.102\\ 
United Kingdom & GBR & 214 & 22 & 229,133 & 0.983 $\pm$ 0.210\\ 
Greece & GRC & 320 & 93 & 7,639 & 0.930 $\pm$ 0.096\\ 
Croatia & HRV & 88 & 40 & 9,672 & 1.085 $\pm$ 0.172\\ 
Hungary & HUN & 143 & 25 & 34,474 & 1.189 $\pm$ 0.238\\ 
Ireland & IRL & 189 & 62 & 4,775 & 1.093 $\pm$ 0.139\\ 
Iceland & ISL & 15 & 12 & 708 & 0.560 $\pm$ 0.162\\ 
Italy & ITA & 400 & 157 & 19,724 & 0.885 $\pm$ 0.071\\ 
Lithuania & LTU & 76 & 32 & 10,654 & 1.007 $\pm$ 0.178\\ 
Latvia & LVA & 75 & 28 & 9,276 & 1.107 $\pm$ 0.209\\ 
Republic of Moldova & MDA & 31 & 23 & 6,609 & 0.570 $\pm$ 0.119\\ 
Macedonia & MKD & 45 & 23 & 11,001 & 0.981 $\pm$ 0.205\\ 
Netherlands & NLD & 69 & 16 & 112,058 & 1.288 $\pm$ 0.322\\ 
Norway & NOR & 105 & 18 & 21,795 & 1.214 $\pm$ 0.286\\ 
Poland & POL & 236 & 160 & 17,390 & 0.903 $\pm$ 0.071\\ 
Portugal & PRT & 139 & 32 & 17,110 & 1.027 $\pm$ 0.182\\ 
Romania & ROU & 522 & 385 & 3,129 & 0.740 $\pm$ 0.038\\ 
Russia & RUS & 622 & 384 & 31,964 & 0.893 $\pm$ 0.046\\ 
Serbia and Montenegro & SCG & 60 & 27 & 38,415 & 1.340 $\pm$ 0.258\\ 
Slovakia & SVK & 88 & 20 & 35,068 & 1.468 $\pm$ 0.328\\ 
Slovenia & SVN & 88 & 32 & 3,273 & 0.730 $\pm$ 0.129\\ 
Sweden & SWE & 168 & 61 & 11,449 & 1.008 $\pm$ 0.129\\ 
Ukraine & UKR & 164 & 107 & 36,515 & 0.833 $\pm$ 0.081\\
\caption{{\bf European countries.} We show the name, the ISO 3166-1 alpha-3
code, the number of cities obtained by the City Local Clustering Algorithm
(CLCA) and the number of those covered by the maximum likelihood power-law fit
defined by the Maximum Likelihood Estimator (MLE) \cite{clauset2009}
(represented by $\dagger$), the lower bound $P_{min}$, and the Zipf's exponent
$\zeta$.}
\label{t_countries_eu}\\
\end{longtable}
\clearpage

\LTcapwidth=\textwidth
\begin{longtable}{c|c|c|c|c|c}
Country & ISO & CLCA cities & CLCA cities$^\dagger$ & $P_{min}$ & $\zeta$\\ \hline
Canada & CAN & 1,135 & 308 & 4,879 & 0.815 $\pm$ 0.046\\ 
Costa Rica & CRI & 14 & 11 & 20,751 & 1.195 $\pm$ 0.360\\ 
Cuba & CUB & 113 & 46 & 34,673 & 1.327 $\pm$ 0.196\\ 
Guatemala & GTM & 25 & 14 & 28,353 & 0.948 $\pm$ 0.253\\ 
Honduras & HND & 236 & 35 & 17,120 & 1.290 $\pm$ 0.218\\ 
Haiti & HTI & 23 & 18 & 21,953 & 0.897 $\pm$ 0.211\\ 
Mexico & MEX & 474 & 284 & 11,992 & 0.726 $\pm$ 0.043\\ 
Nicaragua & NIC & 31 & 28 & 9,802 & 0.821 $\pm$ 0.155\\ 
Panama & PAN & 40 & 12 & 17,717 & 1.089 $\pm$ 0.314\\ 
El Salvador & SLV & 25 & 13 & 21,323 & 0.816 $\pm$ 0.226\\ 
United States & USA & 22,893 & 1,624 & 9,874 & 0.876 $\pm$ 0.022\\
\caption{{\bf North American countries.} We show the name, the ISO 3166-1
alpha-3 code, the number of cities obtained by the City Local
Clustering Algorithm (CLCA) and the number of those covered by the maximum
likelihood power-law fit defined by the Maximum Likelihood Estimator (MLE)
\cite{clauset2009} (represented by $\dagger$), the lower bound $P_{min}$, and
the Zipf's exponent $\zeta$.}
\label{t_countries_na}\\
\end{longtable}
\clearpage

\LTcapwidth=\textwidth
\begin{longtable}{c|c|c|c|c|c}
Country & ISO & CLCA cities & CLCA cities$^\dagger$ & $P_{min}$ & $\zeta$\\ \hline
Australia & AUS & 177 & 145 & 5,332 & 0.788 $\pm$ 0.065\\
Fiji & FJI & 15 & 14 & 936 & 0.807 $\pm$ 0.216\\
Marshall Islands & MHL & 28 & 27 & 44 & 0.760 $\pm$ 0.146\\
New Zealand & NZL & 108 & 79 & 3,077 & 0.776 $\pm$ 0.087\\
Papua New Guinea & PNG & 30 & 13 & 13,828 & 1.479 $\pm$ 0.410\\
\caption{{\bf Oceanian countries.} We show the name, the ISO 3166-1 alpha-3
code, the number of cities obtained by the City Local Clustering Algorithm
(CLCA) and the number of those covered by the maximum likelihood power-law fit
defined by the Maximum Likelihood Estimator (MLE) \cite{clauset2009}
(represented by $\dagger$), the lower bound $P_{min}$, and the Zipf's exponent
$\zeta$.}
\label{t_countries_oc}\\
\end{longtable}
\clearpage

\LTcapwidth=\textwidth
\begin{longtable}{c|c|c|c|c|c}
Country & ISO & CLCA cities & CLCA cities$^\dagger$ & $P_{min}$ & $\zeta$\\ \hline
Argentina & ARG & 749 & 227 & 10,880 & 0.994 $\pm$ 0.066\\
Bolivia & BOL & 83 & 57 & 6,729 & 0.841 $\pm$ 0.111\\
Brazil & BRA & 966 & 613 & 18,555 & 1.057 $\pm$ 0.043\\
Chile & CHL & 59 & 19 & 93,915 & 1.422 $\pm$ 0.326\\
Colombia & COL & 402 & 163 & 12,890 & 0.886 $\pm$ 0.069\\
Ecuador & ECU & 94 & 54 & 12,717 & 0.832 $\pm$ 0.113\\
Peru & PER & 417 & 153 & 8,279 & 0.867 $\pm$ 0.070\\
Paraguay & PRY & 29 & 26 & 4,928 & 0.700 $\pm$ 0.137\\
Uruguay & URY & 79 & 16 & 23,346 & 1.310 $\pm$ 0.327\\
Venezuela & VEN & 81 & 28 & 82,323 & 1.254 $\pm$ 0.237\\
\caption{{\bf South American countries.} We show the name, the ISO 3166-1
alpha-3 code, the number of cities obtained by the City Local Clustering
Algorithm (CLCA) and the number of those covered by the maximum likelihood
power-law fit defined by the Maximum Likelihood Estimator (MLE)
\cite{clauset2009} (represented by $\dagger$), the lower bound $P_{min}$, and
the Zipf's exponent $\zeta$.}
\label{t_countries_sa}\\
\end{longtable}
\clearpage

\LTcapwidth=\textwidth
\begin{longtable}{c|c|c|c|c}
Continent/Globe & CLCA cities & CLCA cities$^\dagger$ & $P_{min}$ & $\zeta$\\ \hline
Africa & 4,860 & 660 & 61,569 & 0.940 $\pm$ 0.037\\ 
Asia & 10,953 & 1,167 & 169,588 & 0.947 $\pm$ 0.028\\ 
Europe & 6,118 & 1,489 & 33,951 & 0.895 $\pm$ 0.023\\ 
Oceania & 180 & 103 & 2,668 & 0.745 $\pm$ 0.073\\ 
N. America & 24,919 & 1,364 & 20,373 & 0.883 $\pm$ 0.024\\ 
S. America & 2,934 & 522 & 39,514 & 0.929 $\pm$ 0.041\\ 
World (except Antarctica) & 50,314 & 8,019 & 35,725 & 0.871 $\pm$ 0.010\\
\caption{{\bf Continents and the entire world.} We show the name, the number of
cities obtained by the City Local Clustering Algorithm (CLCA) and the number of
those covered by the maximum likelihood power-law fit defined by the Maximum
Likelihood Estimator (MLE) \cite{clauset2009} (represented by $\dagger$), the
lower bound $P_{min}$, and the Zipf's exponent $\zeta$.}
\label{t_continents_world}\\
\end{longtable}
\clearpage

\LTcapwidth=\textwidth
\begin{longtable}{c|c|c|c}
CLCA City & Country & CLCA population (people) & CLCA Area ($km^2$)\\ \hline
Alexandria-Cairo-Luxor & Egypt & 63,585,039 & 34,434\\
Dhaka & Bangladesh & 48,419,117 & 26,963\\
Guangzhou-Macau-Hong Kong & China & 44,384,647 & 12,896\\
Tokyo & Japan & 34,318,072 & 9,189\\
Kolkota & India & 28,876,910 & 10,408\\
Patna & India & 28,484,380 & 18,670\\
Xi'an & China & 25,370,875 & 39,736\\
Jakarta-Bekasi-Banten & Indonesia & 23,814,197 & 5,862\\
Hanoi-Hai Phong & Vietnam & 22,480,083 & 19,128\\
New Delhi & India & 22,136,675 & 6,914\\
Seoul & South Korea & 20,318,881 & 3,610\\
Mumbai & India & 18,431,960 & 2,443\\
Manila & Philippines & 17,591,794 & 4,039\\
Mexico City & Mexico & 17,190,725 & 2,845\\
S\~ao Paulo & Brazil & 16,984,627 & 2,840\\
Kyoto-Osaka-Kobe & Japan & 16,398,829 & 4,608\\
New York City & US & 16,364,109 & 4,471\\
Shangai & China & 15,291,143 & 2,529\\
Kochi-Kottayam-Kollam & India & 14,551,809 & 8,091\\
Surabaya-Gresik-Malang & Indonesia & 14,289,547 & 6,891\\
Los Angeles & US & 13,615,610 & 5,167\\
Cirebon-Tegal-Kebumen & Indonesia & 12,758,617 & 6,818\\
Semarang-Klaten-Surakarta & Indonesia & 12,456,408 & 6,418\\
Moscow & Russia & 11,894,034 & 1,448\\
Buenos Aires & Argentina & 11,132,081 & 2,653\\
\caption{{\bf Top 25 cities, by population, in the world.} We emphasize that,
after the top CLCA city (Alexandria-Cairo-Luxor), the 13 next-largest CLCA
cities are in Asia. The largest United Nation (UN) city, Tokyo, is just the 4th
according to our analyses.}
\label{t_world}\\
\end{longtable}

\end{document}